\documentclass[reprint,superscriptaddress,amsmath,amssymb,aps,pra]{revtex4-1} 
\usepackage{graphicx}
\usepackage{dcolumn}
\usepackage{amsmath}
\usepackage{amssymb}

\providecommand{\openone}{\leavevmode\hbox{\small1\kern-3.8pt\normalsize1}}
\graphicspath{{images/}}

\newcommand{\mcE}{\mathcal{E}}

\newcommand{\Eref}[1]{Eq.~(\ref{#1})} 

\begin{document}
\bibliographystyle{apsrev}
\newcommand{\cu}{Universidad Nacional Aut\'onoma de M\'exico, M\'exico D. F. 01000, Mexico}
\newcommand{\icf}{Instituto de Ciencias F\'{\i}sicas, Universidad Nacional Aut\'onoma de M\'exico, Cuernavaca 62210, Mexico}
\newcommand{\ifunam}{Instituto de F\'{\i}sica, \cu}
\newcommand{\iimas}{Instituto de Investigaciones en Matem\'aticas Aplicadas y Sistemas, \cu}
\title{Quantum estimation of unknown parameters}
\author{Esteban Mart\'inez-Vargas}
\affiliation{\iimas}
\author{Carlos Pineda}
\affiliation{\ifunam}
\author{Fran\c{c}ois Leyvraz}
\affiliation{\icf}
\author{Pablo Barberis-Blostein}
\affiliation{\iimas}

\date{\today}

\begin{abstract}
We discuss the problem of finding the best measurement strategy for
estimating the value of a quantum system parameter.
In general the {\em optimum} quantum measurement, in the sense that it maximizes 
the quantum Fisher information and hence allows one to minimize the estimation error,
can only be determined if the value of the parameter is already known.
A modification of the quantum Van Trees inequality, which gives a
lower bound on the error in the estimation of a random parameter, is
proposed. The suggested inequality allows us to assert if a particular
quantum measurement, together with an appropriate estimator, is
optimal.
%
An adaptive strategy to estimate the value of a parameter, based on
our modified inequality, is proposed.
%
%
\end{abstract}
\maketitle
\section{Introduction} 
The problem of determining the value of an unknown parameter $\theta$
from a set of measurements which depend on $\theta$ probabilistically,
has a long history~\cite{helstrom_minimum_1967,braunstein_statistical_1994,Giovannetti2011}.
When one wants to estimate $\theta$, one, in general,
does not measure $\theta$ or even a value in one to one correspondence with it~\cite{escher_quantum_2011}. 
One rather obtains a  variable ${\bf y}$,  
generally vectorial, chosen from 
a probability distribution that depends on $\theta$. The probability 
of obtaining ${\bf y}$ is written as $p({\bf y}| \theta)$.
%
Let us now assume
that we have actually measured ${\bf y}$ and we consider a function
$\hat{\theta}({\bf y})$, which we call the estimator of $\theta$ and
provides a---necessarily imperfect---estimation of the value of
$\theta$.\par

We would now like to distinguish between good and bad estimators, and
to find out when a given estimator is the best possible. We define the
{\em variance\/} of the estimator as follows:
\begin{equation}
\sigma^2\left[
\hat{\theta}({\bf y});\theta
\right]:=\int d{\bf y}\,p({\bf y}|\theta)\left(
\hat{\theta}({\bf y})-\theta
\right)^2.
\label{eq:sigma}
\end{equation}
A fundamental result states that for an {\em
unbiased\/} estimator, that is one which averaged over $p(\bf
y|\theta)$ yields the value $\theta$, the variance is bounded
from below by
\begin{equation}
\sigma^2\left[
\hat{\theta}({\bf y});\theta
\right]\geq\left(
\int d{\bf y}\,  
\Omega[p(\bf y |\theta)]
\right)^{-1}
=I\left(
\theta
\right)^{-1}.
\label{eq:2}
\end{equation}
The operator $\Omega$ acts on functions, taking $\alpha(\theta)$ to
$\Omega[\alpha(\theta)]=(\partial_\theta \ln \alpha)^2 \alpha$.
Here $I(\theta)$ is known as the Fisher information. \Eref{eq:2} is
the so-called Cram\'er--Rao inequality~\cite{Fisher309,*cramer1945mathematical,*Rao1992}. Note that the Fisher
information only depends on the value of $\theta$, as well as, of
course, on the probability distribution $p({\bf y}|\theta)$ and in no
way on the estimator. Thus, if we know the probabilistic
model $p({\bf y}|\theta)$,
this inequality gives a limit to the attainable variance for any
unbiased estimator $\hat{\theta}({\bf y})$. It therefore states, among
other things, when an estimator is optimal.

However, the inequality is of limited use if one does not know a way of
finding estimators which saturate this bound, at least approximately.
Such a procedure exists for a broad class of cases and 
leads to the so-called maximum likelihood estimator.
Note that, if the vector $\bf y$
consists of $n$ independent measurements of a quantity $x$
having probability distribution $p(x|\theta)$, then (\ref{eq:2})
immediately leads to a lower bound of $1/n$ for the variance.
Therefore, in this case, it is impossible to obtain estimators that
are systematically closer to $\theta$ than $n^{-1/2}$.
This rate of convergence is a central and very general feature 
in classical statistics. Using quantum mechanics, one may do better~\cite{PhysRevLett.71.1355}, but this will
not be the point of this paper.

Now let us consider the same problem for a quantum mechanical system.
We consider a density matrix $\rho(\theta)$ depending in a known
manner on an unknown parameter $\theta$, which one wishes to determine on the
basis of measurements performed on the system;
for example in an interferometry experiment we may want to 
estimate a phase shift due to the presence of
a crystal.
The crucial difference
with the classical case has to do with the issue of measurement, which
is more complex in quantum systems. Indeed, {\it the choice of the
observable to be measured determines which aspects of the quantum
system will appear}.

We must therefore first define clearly what we mean by measurement. We
shall take a somewhat more general definition than that commonly used
in textbooks: We define a set of (not necessarily commuting)
positive operators $\{ E_\xi \}$ indexed by a parameter $\xi$ to be a {\em
positive operator-valued measure\/} (POVM) if it satisfies
$\sum_\xi E_\xi= \openone$.
The outcome of applying such a POVM to a given density matrix $\rho$
is one of the parameter values $\xi$, with the probability
$p(\xi)=\mbox{\rm Tr}(\rho E_\xi)$.
Note that, if $\{ E_\xi \}$ is a set of commuting projectors satisfying
the aforementioned identity, the procedure reduces to the traditional
quantum mechanical prescription.
Advantages of considering POVM's
are described in~\cite{Nielsen:2011:QCQ:1972505}.

Once a POVM has been determined, the problem is reduced to a classical
one: if the state of the system is given by $\rho(\theta)$ and the POVM is given by
$\{ E_\xi \}$, the probabilities of obtaining $\xi$, given the value of $\theta$, are given by:
\begin{equation}
p(\xi|\theta)=\mbox{\rm Tr}\left[
\rho(\theta)E_\xi
\right].
\label{eq:5}
\end{equation}
It might now seem that everything is solved and reduced to finding a
POVM which maximizes the Fisher information of $p(\xi|\theta)$. The
difficulty that arises is the following: this optimal POVM generally
depends on the actual value of
$\theta$~\cite{barndorff-nielsen_fisher_2000}, which is, however,
always assumed to be unknown. {\it Indeed, the very problem one is trying to
solve is that of estimating $\theta$}.

\section{The Van Trees Information} 
So how can one solve this conundrum? If several copies of the system are
available, solutions of this conundrum can be found in the limit where the
quantum Fisher information does not depend on the value of the
parameter~\cite{Jaryzna:2015} or using adaptive
measurements~\cite{fujiwara_strong_2006}. These strategies depend on i) having
access to several copies of the system and, for the adaptive measurement
solution, ii) the possibility of changing, prior to the measurement of each copy,
the POVM to be used. What is the solution if i) and/or ii) are not satisfied?.

We propose the following: Let us start by introducing a probability
distribution $\lambda(\theta)$, according to which we assume the
values of $\theta$ are distributed. We do not, of course, necessarily
have such information, but we surely have some idea of the range of
values the parameter $\theta$ is liable to take, and we may at least
incorporate such knowledge into $\lambda(\theta)$. Instead of the
variance defined in (\ref{eq:2}), we define the quality of the
estimator through the quantity
\begin{equation}
\sigma^2[\hat{\theta}({\bf y}); \lambda]:=\int d{\bf y}\,d\theta\,\lambda(\theta)\,p({\bf y}|\theta)\left(
\hat{\theta}({\bf y})-\theta
\right)^2,
\label{eq:6}
\end{equation}
corresponding to the estimator variance {\em averaged\/} over the
distribution of $\theta$. In such cases, a generalization of the
Cram\'er--Rao inequality has been derived, the so-called Van Trees
inequality~\cite{gill_applications_1995}:
\begin{equation}
\sigma^2[\hat{\theta}({\bf y});\lambda]\geq Z(\lambda)^{-1}\, .
\label{eq:vantressinequality}
\end{equation}
with 
\begin{equation}
Z(\lambda) \equiv
\int d{\bf y}\,d\theta
\big\{
  \Omega[p({\bf y}|\theta)] + \Omega[\lambda(\theta)]
\big\}
p({\bf y}|\theta)\lambda(\theta) .
\label{eq:vantressdefinition}
\end{equation}
In words, the lower bound on the average variance is the inverse of
the average over the joint distribution of ${\bf y}$ and $\theta$ of
the sum of the Fisher informations of $p({\bf y}|\theta)$ and
$\lambda(\theta)$. We will call $Z(\lambda)$ the generalized Fisher
information~\cite{paris_quantum_2009}. The Cram\'er-Rao bound~(\ref{eq:2}) can
be obtained from~(\ref{eq:vantressdefinition}) when the prior probability
distribution $\lambda(\theta)$ is constant and the Fisher information does not
depends on the parameter. If $\lambda(\theta)$ is a Dirac delta, $Z(\lambda)$
diverges and the error for the optimal strategy is zero. This makes sense
because the prior knowledge of the parameter gives complete knowledge of it.

For a quantum mechanical systems described by
a density matrix $\rho(\theta)$ depending in a known manner on an
unknown parameter, we follow an approach analog to what was done
above. Using a POVM we construct a probabilistic model through
(\ref{eq:5}), then we can use (\ref{eq:vantressinequality}) to find the optimal
estimator. The POVM that maximizes $Z(\lambda)$ --the optimal POVM--
together with the estimator that saturates \Eref{eq:vantressinequality} minimizes the
error~(\ref{eq:6}); they will be called the optimal measurement
strategy. Let $\mcE=\{E_\xi\}$ represents the set of all POVMs acting on
the quantum system. 
If we define the quantum Van Trees information as
\begin{align}
Z_Q(\lambda) &= \max_{\{E_\xi\}} \int d{\xi}\,d\theta
   \big\{ \Omega[ p_{E}({\xi}|\theta) ] + \Omega[\lambda(\theta)] \big\}
p_{E}({\xi}|\theta)\lambda(\theta) \nonumber \\
&=\max_{\mcE} Z(\lambda)\, ,
\label{eq:VTI}
\end{align}
a Cram\'er-Rao type equation for the quantum case
can be written as
\begin{equation}
  \label{eq:8}
\sigma^2[\hat{\theta}({\xi});\lambda] \geq \big[ Z_Q(\lambda) \big]^{-1}.
\end{equation}
The POVM and estimator that saturates it constitute the optimal
strategy for estimating $\theta$. 

Inequality~(\ref{eq:8}) resembles the quantum Van Trees inequality,
which can be stated defining the
generalized quantum Fisher information,
\begin{equation}
V_Q(\lambda) = \int d{\xi}\,d\theta \max_{\{E_\xi\}} 
   \big\{ \Omega[ p_{E}({\xi}|\theta) ] + \Omega[\lambda(\theta)] \big\}
p_{E}({\xi}|\theta)\lambda(\theta) \, ,\nonumber
\label{}
\end{equation}
and takes the simple form \cite{paris_quantum_2009}
\begin{equation}
  \label{eq:QVTI}
\sigma^2[\hat{\theta}({\xi});\lambda]\geq \left(V_Q(\lambda)\right)^{-1}\, .
\end{equation}
Notice that the maximization over all POVMs is taken inside the
integral whereas in~(\ref{eq:VTI}) it is taken outside the integral.
Since the POVM that maximizes the Fisher information depends on the
value of the parameter to estimate, there is not, in general, a single
POVM that saturates the quantum Van Trees inequality; in this case the
inequality is useless to find the best POVM for parameter estimation.
By construction $V_Q\geq Z_Q$. When $V_Q>Z_Q$, the Van Trees inequality
predicts smaller errors for the optimal measurement strategy; it can then be
argued that the Van Trees inequality~(\ref{eq:QVTI}) gives better results, but
as shown above, the POVM that saturates it does not exist and {\it no strategy
exists to achieve the smaller error predicted by~(\ref{eq:QVTI}).}

If only one measurement over a quantum state $\rho(\theta)$ is
allowed, and we codify what we know about the parameter in the
probability distribution $\lambda(\theta)$, the
inequality~(\ref{eq:8}) tells us that in order to minimize the
error~(\ref{eq:6}) we should use the quantum measurement that
maximizes $Z(\lambda)$. This POVM does not depends on the parameter we
want to estimate and thus, {\it this way of deciding how to measure
  the quantum state solves the conundrum discussed above}. In our
proposal the optimal measurement strategy depends on what we already
know about the parameter, i.e., the {\it a priori} probability
distribution, and not on what will be estimated {\it after} the
measurement. Once the optimal POVM has been found, the problem reduces to a
classical one and the mean of the posterior probability distribution minimizes
the risk~(\ref{eq:6})~\cite{Blume-Kohout:2010}.
%
Note that Van Trees inequalities do not require unbiased estimators, and one thus
expects them to provide better error bounds than the usual Cram\'er-Rao
inequalities~\cite{1605.03799}. This fact underlines the importance of
\Eref{eq:8}.

\section{Measuring several copies of the system} 
Suppose we have a machine that generates $n$ copies of the quantum
state $\rho(\theta)$ and we want to estimate $\theta$ with the
smallest error. We consider three different strategies to
  minimize the error in the estimation, each of them depending on the
  type of measurement that can be done on the $n$ copies. The
conundrum described in the introduction does not appear because the
POVMs that minimizes the error in each of the strategies to be
described do not depend on the value of the parameter we want to
estimate.

\subsection{Any possible measurement on the $n$ copies}

If we can make any possible quantum measurement---including collective
ones---over the $n$ copies, the POVM that maximizes the generalized
Fisher information, see Eq.~(\ref{eq:VTI}), corresponds to the
mea\-sure\-ment that gives the smallest error given the prior
information. This is the best case scenario.

\subsection{Independent measurement on each copy} 
\label{sc:independent_measurement}
Let us think of a more plausible situation: The states are provided
in such a way that collective measurements over two or more copies
are not possible. For example, the machine gives the states
sequentially and the time it needs to create the next copy is larger
than the decoherence time of the state, so that there are never two
or more copies of the same state available to make a collective
measurement; then only $n$ independent measurements, each on one of
the $n$ copies of the system, can be performed. If the same
measurement is going to be performed in each copy, we can use the
additive property of the Fisher Information for independent
measurements to calculate the quantum Van Trees information for $n$
independent measurements performed with the same POVM,
\begin{align}
Z^{I}_Q(\lambda) = \max_{\{E_\xi\}} \int d{\xi}\,d\theta
& \big\{ n\times\Omega[ p_{E}({\xi}|\theta) ] \nonumber\\
& + \Omega[\lambda(\theta)] \big\}
p_{E}({\xi}|\theta)\lambda(\theta) \, .
\label{eq:VTI_independent}
\end{align}
The maximization is done over all the POVMs acting on one copy of the
system; the number of measurements performed appears as a factor
multiplying the first summand of the integral in the previous
equation. The Cram\'er-Rao bound reduces to
\begin{equation}
\sigma^2[\hat{\theta}({\xi});\lambda] \geq \big[ Z^{I}_Q(\lambda)
  \big]^{-1}\, .
\end{equation}

For the case of performing the same measurement over several copies
of the system, the risk~(\ref{eq:6}) is minimized by the measurement
that maximizes the integral in~(\ref{eq:VTI_independent}).

\subsection{Adaptive measurements} 

Lets assume now a setup similar to the previous one but allowing the
use of the outcomes of previous measurements to choose how to measure
the next copy. We model this situation by choosing $n$ different
individual quantum measurements, one for each copy of the state. The
quality of this adaptive estimation process is given by~(\ref{eq:6})
with ${\bf y}=(\xi_1,\xi_2,\cdots,\xi_{n})$ the outcomes of measuring
$n$ copies of the quantum state $\rho(\theta)$. We now discuss
possible ways to choose this quantum measurement.


In~\cite{fujiwara_strong_2006} the following adaptive scheme is
proposed: First, arbitrarily guess a value for the parameter $\theta$.
Let us call it $\theta_1$. Then, use the quantum Fisher information to
choose the POVM optimized for $\theta_1$, make a measurement with
outcome $\xi$ and estimate $\theta$ using the maximum likelihood
function. We call the second estimation $\theta_2$. Then, repeat the
procedure using the POVM optimized for $\theta_2$. The procedure is
repeated $n$ times. It is shown that in the limit of $n$ going to
infinity the procedure saturates the Cram\'er-Rao inequality and is,
in this sense, optimal. Nevertheless, preparing several copies of a
quantum state and measuring them can be expensive, difficult, or time
consuming; when we are in this situation strategies that are designed
to get smaller estimation errors when $n$ is small are of interest. 

We propose an adaptive method similar to the adaptive scheme proposed
in~\cite{fujiwara_strong_2006} but using the quantum Van Trees information
instead of the quantum Fisher information: We use the 
{\it a priori} knowledge distribution $\lambda(\theta)$ to obtain a
POVM~$\{E_\xi\}$~which maximizes $Z(\lambda)$. A measurement is carried out with
this POVM obtaining $\xi$ as its result with a probability
$p(\xi|\theta)$ given by~(\ref{eq:5}). Then, we use the Bayes rule to
obtain a new probability distribution
\begin{equation}
  \label{eq:bayes_rule}
  \lambda_1(\theta)=\frac{p(\xi|\theta)\lambda(\theta)}{\int p(\xi|\theta^\prime)\lambda(\theta^\prime)
  d\theta^\prime}.
\end{equation}
Afterwards, we calculate $Z_Q(\lambda)$ using $\lambda_1(\theta)$ as our
new {\it a priori} knowledge distribution. A new POVM is obtained and the
process can be iterated from here on $n$ times.
Our method
has the advantage that all the information we know about the parameter
enters in the determination of the optimal POVM. When the number of measurements goes
to infinity both methods predict the same result. We now show an example
where for a small number of measurements our approach is better than
the one in~\cite{fujiwara_strong_2006}.
\section{Example}\label{sc:example} 
\begin{figure} \centering 
	\vspace{5 mm}
  \includegraphics{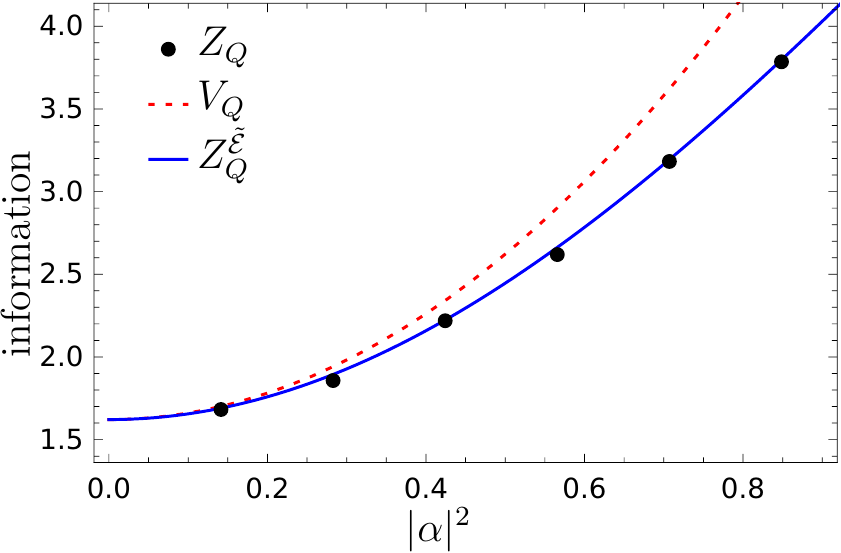}
  \caption{The blue continuous curve is our analytical approximation
    to the quantum Van Trees information for phase estimation as a
    function of the coherent state $|\alpha\rangle$. The {\it a priori}
    probability distribution is a Gaussian centered in zero with width
    $\sigma=\pi/4$. The black dots
    are calculated numerically. The red dashed curve is the
    generalized quantum information. }
  \label{fig:vantreesrandomcoherent}
\end{figure} 
We study the quantum Van Trees information and apply the adaptive
estimation process outlined above for the problem of phase estimation with an
initial coherent state~\cite{escher_quantum_2011}.  Specifically, we consider
estimating $\theta$ from  the state defined by
\begin{equation}
|\alpha(\theta) \rangle =
e^{i\hat{n}\theta}|\alpha\rangle
\label{eq:defalpha}
\end{equation}
where $|\alpha\rangle$ is the coherent state corresponding  
to the complex parameter $\alpha$.
The {\it a priori} distribution $\lambda(\theta)$ is assumed to be a
Gaussian centered at 0 and of width $\sigma$
(we ignore the effect of periodicity in $\theta$).
We shall present an analytical approach,
which involves a restriction over the considered POVM's, and a numerical one,
which will provide an independent check to validate the analytical approach. 


The problem of maximizing Fisher information $I(\theta)$ for the 
phase estimation problem, with pure states, has been solved in
\cite{escher_quantum_2011}. The
corresponding POVM belongs to the family characterized by the set of operators
\begin{equation} 
\label{eq:povms}
\tilde \mcE= \left\{
|\alpha(\epsilon)\rangle\langle\alpha(\epsilon)|, \,
\openone-|\alpha(\epsilon)\rangle\langle\alpha(\epsilon)|
\right\}_{\epsilon\in[0,2\pi)}.
\end{equation} 
Here $|\alpha(\epsilon)\rangle$ is the state defined by (\ref{eq:defalpha}). 
In particular, the POVM that maximizes $I$ is obtained setting $\epsilon$ to
the actual value of the parameter to estimate. This leads to  a value of
$4|\alpha|^2$ for the Fisher information, independent of $\theta$, which 
in turn means that $V_Q = 4|\alpha|^2 + \sigma^{-2}$. 

To calculate the quantum Van Trees information,
the maximization in \Eref{eq:VTI} should be done over all POVMs.
However, to achieve analytic results, we shall restrict the optimization to the
family \Eref{eq:povms}, inspired by the similarity of the Fisher information
problem and the current one. Coincidence with the numeric method will justify this massive
simplification. 
%
%
%
%

If we define $p(\theta,\epsilon) =|\langle\alpha(\theta)|
\alpha(\epsilon)\rangle|^ 2$, the conditional probability distribution reads
$p(1|\theta) = p(\theta,\epsilon)$ and $p(2|\theta) = 1-p(\theta,\epsilon)$.
The Fisher information for a POVM with parameter $\epsilon$ is then
\begin{equation}
  F(\theta,\epsilon)
     =\frac{1}{p(\theta,\epsilon)(1-p(\theta,\epsilon))}
         \left[\frac{dp(\theta,\epsilon)}{d\theta}\right]^2\, .
\label{ec:fisherclas}
\end{equation} 
Notice that $\theta$ is the parameter we want to estimate and
$\epsilon$ is the label that enumerates different POVMs.
We restrict the maximization in (\ref{eq:VTI}) to the subset
$\tilde{\cal E}$ and obtain
\begin{multline} \label{eq:zqe_example}
Z_Q^{\tilde \mcE} \equiv \max_{\tilde \mcE}\left[\int
    \frac{1}{p(\theta,\epsilon)(1-p(\theta,\epsilon))}
     \left[\frac{dp(\theta,\epsilon)}{d\theta}\right]^2\lambda(\theta)d\theta\right]
  \\ +
    \int\left(\frac{\partial\ln\lambda(\theta)}{\partial\theta}\right)^2
      \lambda(\theta)d\theta.
\end{multline} 
Since the maximization is done over a proper subset of all 
POVMs, $Z_Q^{\tilde \mcE}\leq Z_Q$. 
%
With the additional assumption 
$|\alpha|^2\ll 1$ we find, after a lengthy but straightforward calculation
\cite{long:paper},
\begin{equation}\label{eq:qvti_analitical}
Z_Q^{\tilde \mcE} \approx
2|\alpha|^2(e^{-\frac{\sigma^2}{2}}+1)+\frac{1}{\sigma^2} .
\end{equation}
The first term of the sum coincides with the optimized Fisher
information when $\sigma\rightarrow 0$, as expected. 
For large $\sigma$,  $Z_Q^{\tilde \mcE}$ is smaller
than the quantum Fisher information, as we are estimating a parameter about
which  we have little information. 

%

To numerically estimate  $Z_Q$ we first truncate the infinite dimensional
Hilbert space, in which \Eref{eq:defalpha} lives, to a finite dimension, larger
than $|\alpha|$.
Next, we recall that any POVM is equivalent to a projective measurement in a
larger Hilbert space~\cite{Nielsen:2011:QCQ:1972505}. 
We shall vary the dimension of such an enlarged Hilbert space until the value 
of the Van Trees information no longer changes up to the first two significant
digits.

Notice that the columns of any unitary matrix defines an orthonormal basis,
whose elements define a projective measurement (up to arbitrary values of the
measurement outcome).
%
We select the aforementioned unitary matrix from the Gaussian
unitary ensemble~\cite{balian}, which guarantees a uniform exploration of all
orthonormal bases. Such an ensemble is composed of all unitary matrices, and is
weighted by its Haar measure.  
That way we obtained a POVM in the original space, and thus a particular 
value of the integral within \Eref{eq:VTI}. Repeating the proceedure 
many times, we approach the value that maximizes $Z(\lambda)$. 
We add two remarks. First, 
the value of
the dimension of the truncated Hilbert space was varied, for fixed $\alpha$,
until the correction was smaller than could be noticed by visual inspection of
the plot. Second, a downhill method to optimize the orthonormal basis in the
enlarged Hilbert space, to fine tune the POVM, was used. However, this
procedure only provided a small advantage, suggesting that the landscape of this
very large dimensional space had no small and deep depressions.

In Fig.~(\ref{fig:vantreesrandomcoherent}) it can be seen that our analytical
approximation \Eref{eq:zqe_example} gives very good agreement with the
numerical simulation, suggesting that indeed $Z_Q^{{\tilde \mcE}}\approx Z_Q$. In that
figure we also show that the quantum Van Trees information is
significantly smaller than
$V_Q$; {\it this means that there does not exist a single POVM that saturates the
quantum Van Trees inequality~(\ref{eq:QVTI})}, and thus an optimal strategy must 
come from the  optimization as performed in $Z_Q$ and not $V_Q$. \par
Now we discuss the  adaptive method proposed in this paper to 
estimate the parameter, assuming that one has several 
copies of $\rho(\theta)$.
We shall compare our results 
with the method proposed in~\cite{fujiwara_strong_2006}. For both
methods there are $2^n$ possible outcomes after $n$ binary measurement; its
probability distribution $p({\bf y}|\theta)$ and the estimator depend on
the method used.

We assume that the {\it a priori} probability distribution is flat,
$\lambda_0(\theta)=1/2\pi$. This is the worst case scenario: We only
know that the parameter to be measured is an angle between $0$ and
$2\pi$.
For each possible value of the parameter to be estimated, $\theta_r$,
we simulate both adaptive methods. After $n$ measurements and assuming
that the optimal estimator is used, the smallest error for the quantum
Fisher adaptative method for $\theta_r$ is
$\sigma^2_n(\theta_r)=[I_n(\theta_r)]^{-1}$, where $I_n(\theta_r)$ is the Fisher
information for the probability distribution of the $2^n$ possible
outcomes. Then, the mean error for the quantum fisher
information adaptive method is
\begin{equation}\label{eq:cost_fisher}
  \overline{\sigma^2}_{\rm Fisher}(n)=\int_{-\pi}^\pi\lambda_0(\theta)\sigma^2_n(\theta_r)d\theta_r=\frac{1}{2\pi}\int_{-\pi}^\pi\frac{1}{I_n(\theta_r)}d\theta_r\, .
\end{equation}
After $n$ measurements the smallest error for the quantum Van Trees information
adaptive method is $\sigma^2_{\rm VanTrees}(n)=1/Z_{Q_n}(\lambda_{n-1})$, where
$\lambda_{n-1}$ is the prior distribution probability for the parameter to
estimate after $n-1$ measurements, and $Z_{Q_n}(\lambda_{n-1})$ is the quantum
Van Trees information for that prior distribution probability.

\begin{figure}
\centering
  \includegraphics[width=0.5\textwidth]{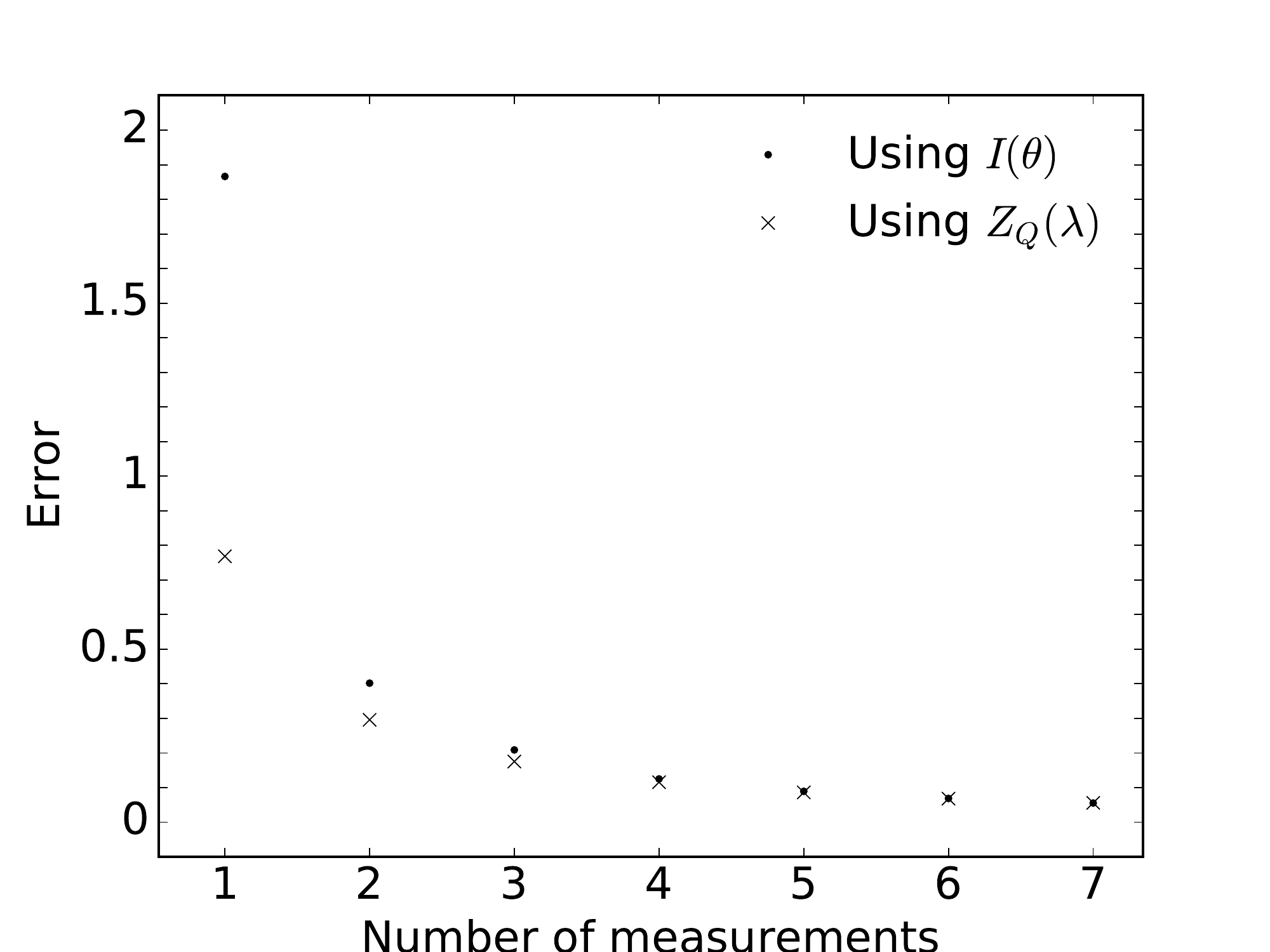}
  \caption{Comparison of the smallest mean error predicted by the
      Fisher Information (Eq.~(\ref{eq:cost_fisher})), with the
      smallest mean error predicted by the quantum Van Trees
      information, for two adaptive quantum estimations schemes. The
    points use the quantum Fisher information as a tool to choose the
    POVM to be used in the next measurement. The crosses use the
    quantum Van Trees information to choose the POVM to be used in the
    next measurement; with this scheme we obtain smaller errors.}
  \label{fig:recursive}
\end{figure}

In Fig.~\ref{fig:recursive} we compare $\overline{\sigma^2}_{\rm Fisher}(n)$
with $\sigma^2_{\rm VanTrees}(n)$; it can be seen that the average estimation
error is smaller using the quantum Van Trees information adaptive
method than the quantum Fisher adaptive method. As expected, both methods tend
to give
the same error as the number of measurements increases.

If the adaptive step cannot be implemented, the same measurement is performed
in all the copies. Using the additive property for the Fisher information for
independent measurements (see Sec.~\ref{sc:independent_measurement}), we get
that $\overline{\sigma^2}_{\rm Fisher}(n)\gtrapprox 1.9/n$ and $\sigma^2_{\rm
VanTrees}(n)\gtrapprox 0.8/n$; for both cases the error scales the same with
respect to the number of measurements, but if we use inequality~(\ref{eq:8}) to
choose the POVM to be implemented, the error in the estimation can be smaller
by more than a factor of two.


\section{Conclusions} 
\label{sc:conclusions}

Assume we have a quantum state that depends on an unknown parameter
chosen from a known distribution. A central problem in quantum
metrology is to determine as accurately as possible the parameter from
measurements on the state. It is customary to use the quantum Fisher
information to find the optimal measurement, which, in general,
depends on the unknown value of the parameter we want to estimate.
Using an inequality proposed in this paper this problem is solved. The
inequality bounds the error in determining the parameter and depends
on its prior distribution. This bound can thus be used to find a
quantum measurement whose results applied to the appropriate estimator
gives the minimum error. The most important application of this
approach consists of determining the optimal way to use whatever {\em
  a priori\/} information is available in the best possible way. This
is particularly important if we can only perform one, or a very small,
number of measurements. We propose an adaptive quantum estimation
scheme, based on the inequality, that can be used when several copies
of the system are available but collective measurements are not
possible.

%

\section{Acknowledgments} 
We acknowledge support from UNAM-PAPIIT Grants No. IN111015, No.
IN114014, No. IN103714 and CONACyT Grant No. 254515. We also
acknowledge support from the high performance computer laboratory
(LUCAR) at IIMAS, as well as useful discussions with Carlton M.
Caves, Mankei Tsang, and Luiz Davidovich.
\bibliography{papervantrees}
\end{document}